\documentclass[aps,pra,twocolumn,showkeys,showpacs,a4paper,nobibnotes,superscriptaddress]{revtex4-2}

\usepackage{graphicx}
\usepackage{amsmath}
\usepackage[colorlinks,citecolor=blue,linkcolor=blue,urlcolor=blue]{hyperref}
\usepackage{soul}

\usepackage{multirow}
\usepackage{booktabs}
\usepackage{tabularx}

\graphicspath{{figures/pdf/}}

\begin{document}

\title{Direct evidence for projectile electronic structure effects in slow multielectron capture collisions}

\author{Akash Srivastav}
\email[]{akash.srivastav@students.iiserpune.ac.in}
\affiliation{Indian Institute of Science Education and Research Pune, Homi Bhabha Road, Pune~411008, India}
\author{Sumit Srivastav}
\affiliation{Indian Institute of Science Education and Research Pune, Homi Bhabha Road, Pune~411008, India}
\affiliation{Universit\'{e} Caen Normandie, ENSICAEN, CNRS, CEA, Normandie Univ., CIMAP UMR6252, F-14000 Caen, France}
\author{Bhas Bapat}
\affiliation{Indian Institute of Science Education and Research Pune, Homi Bhabha Road, Pune~411008, India}

\date{\the\day.\the\month.\the\year}

\begin{abstract}

We investigate the role of the electronic structure of the projectile in ionization and subsequent fragmentation of CO$_2$ induced by multielectron capture in collisions at 0.31~a.u. impact velocity. Focusing on the \mbox{$\text{CO}_2^{3+} \rightarrow \text{O}^+:\text{C}^+:\text{O}^+$} break-up channel as a representative channel, we report kinetic energy release distributions (KERDs) for collisions with equi-velocity N$^{q+}$ and O$^{q+}$ projectiles. We consider two complementary categories of measurements. In the first category, in which different projectiles of the same charge are considered, we find that KERDs obtained with N$^{q+}$ and O$^{q+}$ impact ($q=4,6$) are broadly similar, but they differ significantly from the earlier reported KERD with Ar$^{q+}$ impact. In the second category, pronounced differences are observed between the KERDs obtained with isoelectronic N$^{q+}$ and O$^{(q+1)+}$ ($q=3,5,7$) projectiles. These results provide direct evidence that projectile electronic structure plays a critical role in multielectron capture collisions. 

\end{abstract}


\maketitle


Investigations of collisions between highly charged ions (HCIs) and molecules have attracted sustained attention within the atomic and molecular physics community over the past several decades. Considerable effort has been devoted to the study of the ionization mechanisms and energy transfer to the target, and their ensuing role in the dissociative ionization of molecules. An extensive body of literature addressing the ionization mechanisms and subsequent bond break-up dynamics of the resulting multiply charged molecular ions exists \cite{Luna_2007,Neumann_2010,Wang_2015,Khan_2015,Luna_2016,Sumit_2021,Sumit_2022,Rajput_2018,Severt_2024,Akash_2026_a,Akash_2026_b}. 

In slow collisions \mbox{($v < 1$~a.u.)}, multielectron capture is the dominant ionization mechanism \cite{Vancura_1994,Wu_1995,Wells_2005,Mawhorter_2007,Luna_2016,Wu_2025}. Frameworks, such as the Bohr-Lindhard model \cite{Bohr_1954} and the extended classical over-the-barrier model (ECOBM) \cite{Barany_1985,Niehaus_1986}, treat the projectile as a positively charged Coulomb center which captures electrons from the target, implying that the collision dynamics are governed primarily by the projectile charge $q$ and velocity $v$. In this regime, it is generally understood that, at a fixed collision velocity, increasing $q$ leads to progressively gentler collisions \cite{Folkerts_1996,Khan_2021}. This is inferred from the kinetic energy release (KER: defined as the sum of the fragment kinetic energies) and its distribution (KERD), where increasing $q$ progressively leads to increased propensity of populating low-lying states of the molecular ion.  As a prototypical system for understanding energy deposition, the \mbox{O$^+$:\,C$^+$:\,O$^+$} fragmentation channel of CO$_2^{3+}$ has been extensively investigated using a variety of projectiles. The KERD for this channel is usually dominated by a low-KER peak, indicating preferential population of low-lying electronic states, although the relative intensities of the different features in the KERD vary with the projectile \cite{Neumann_2010,Sumit_2021,Sumit_2022,Kumar_2024,Akash_2026_a,Akash_2026_b}.  However, investigations using wider variety of projectiles have revealed the inadequacy of simple charge and velocity scaling description of the fragmentation dynamics \cite{Akash_2026_a,Kumar_2024,Sumit_2021}. In our recent work on slow Ar$^{q+}$ impact, we reported KERDs for this break-up channel and observed non-systematic variations in the population of high-lying states as a function of $q$. Based on these observations, we argued that the projectile electronic structure, rather than its charge alone, must be considered to understand multielectron capture induced fragmentation of molecules \cite{Akash_2026_a}. However, that conclusion was inferred only indirectly through variations with projectile charge. A direct and systematic assessment requires comparisons in which the projectile charge state and collision velocity are held fixed while varying the projectile species.

In the present work, we implement precisely such a strategy. We investigate the \mbox{O$^+$:\,C$^+$:\,O$^+$} fragmentation channel of CO$_2^{3+}$ produced in slow collisions with N$^{q+}$ and O$^{q+}$ projectiles, with different values of $q$ but the same velocity \mbox{($v = 0.31$~a.u.)}. We also extend the comparison to the earlier results reported for Ar$^{q+}$ projectiles at nearly the same velocity \cite{Akash_2026_a}. N$^{q+}$ and O$^{q+}$ projectiles have far fewer electrons that Ar$^{q+}$ for the same $q$, so the energy landscape available for capture of electrons in the former set is very different from that in the latter.  Thus a comparison across these two sets is well suited for investigating projectile-species effects.  Furthermore, since N and O differ in $Z$ by unity, they provide a platform for investigations with isoelectronic projectiles with nearly the same charge, i.e.\ N$^{q+}$ and O$^{(q+1)+}$, which we carry out. Based on an energy deposition perspective, only minor differences would be expected between such systems, and any variations should follow a simple charge-scaling behavior. 
Deviations from these expectations will thus be direct experimental evidence for the influence of projectile electronic structure on multielectron capture and the ensuing fragmentation dynamics. As we will see, the variations in KERDs and hence the collision dynamics are far too severe to be explained based on projectile charge differences alone. A comprehensive understanding will require incorporation of projectile-specific electronic structure within a quasi-molecular framework, where state-dependent couplings and capture pathways govern how high-lying molecular states get populated and decay.


Experiments were performed using the electron beam ion source facility at IISER Pune \cite{Bapat_2020}. A multihit-capable ion momentum spectrometer (IMS), operated under Wiley-McLaren space-focusing conditions \cite{Wiley_1955}, was used to study the fragmentation dynamics of molecular ions. Post-collision charge-state analysis of the projectile is possible using a cylindrical deflector analyzer (CDA) mounted downstream of the IMS, however this feature was not used for the current study.  Further details of the IMS and CDA are available elsewhere \cite{Vandana_2006,Sumit_2022_CDA}.

The projectile beam from the EBIS intersected an effusive CO$_2$ gas jet at the center of the IMS.  Typical beam currents were $\approx 10$ pA, and the background pressure was maintained at a few $10^{-7}$ mbar, ensuring low accidental-coincidence rates and single-collision conditions.  Recoil and fragment ions produced in the collisions were guided to a position-sensitive ion detector by a uniform extraction field of 60~V/cm, applied perpendicular to both the projectile beam and the gas jet. The same field directed the ejected electrons to a channeltron detector mounted opposite the ion detector.  Fragment and recoil ions were detected in coincidence with ejected electrons. As is well established, in the present collision regime, the electrons used as event start signal predominantly originate from autoionization of the scattered projectile \cite{Cocke_1981,Simcic_2010}. The present measurements involved N$^{q+}$ \mbox{($3\le q\le 7$)} and O$^{q+}$ \mbox{($4\le q\le 8$)} projectiles colliding with CO$_2$ molecules at identical collision velocities of 0.31~a.u. Earlier measurements with Ar$^{q+}$ at the same velocity were carried out under nearly identical conditions.
 


\begin{figure}
	\centering
	\includegraphics[width=0.9\columnwidth]{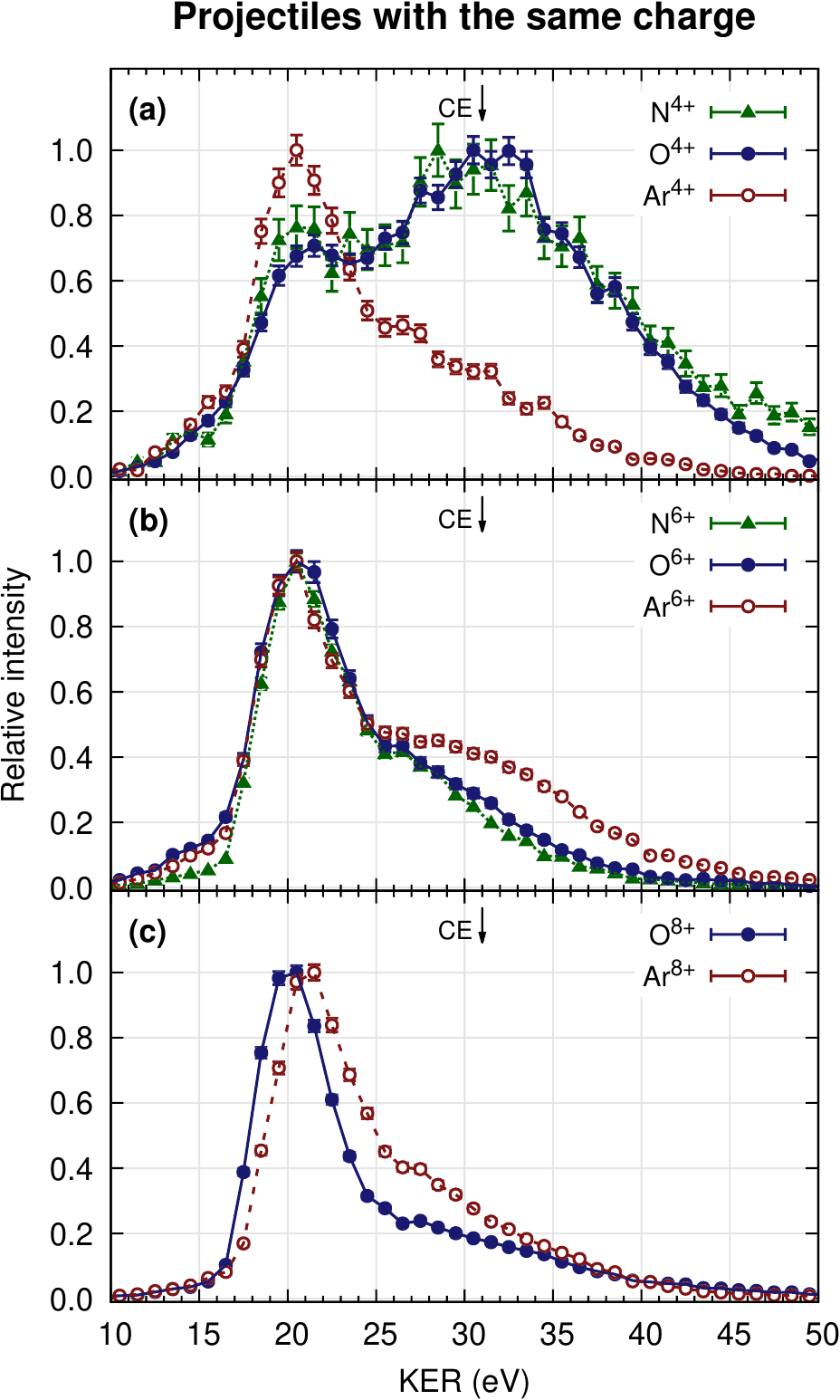}
	\caption{KERDs (normalised to the highest count) for the \mbox{O$^+$:\,C$^+$:\,O$^+$} break-up channel of CO$_2^{3+}$, for various projectiles at nearly identical collision velocities: $v=0.27$~a.u.\ for Ar$^{4+}$ and $v=0.31$~a.u.\ for all other projectiles. In each panel, the KERDs are compared for projectiles having the same charge. The data for Ar$^{q+}$ projectiles are taken from Ref.~\cite{Akash_2026_a}. The arrows indicate the KER expected from a simple point-charge CE model. Error bars on the experimental data represent statistical uncertainties.}
	\label{figure1}
\end{figure}

We first focus on KERDs obtained with projectiles having the same charge but different electronic configurations: N$^{q+}$, O$^{q+}$, and Ar$^{q+}$. These are shown in Fig.~\ref{figure1}. The arrows indicate the KER expected from a simple point-charge Coulomb explosion (CE) model. The data for Ar$^{q+}$ projectiles are taken from our earlier report \cite{Akash_2026_a}. Except for N$^{4+}$ and O$^{4+}$, the KERDs for all projectiles exhibit a prominent peak around 21~eV, accompanied by a broad shoulder above $\sim$25~eV and an extended high-energy tail reaching up to 50~eV. The peak around 21~eV was previously attributed to the low-lying $^{2,4}\Pi_\text{g,u}$, $^{2,4}\Sigma_\text{g}^+$, and $^{2,4}\Delta_\text{g}$ electronic states of CO$_2^{3+}$, while the structures near \mbox{27--31}~eV in the high-KER shoulder were assigned to the $^{6}\Pi_\text{g,u}$, $^{6}\Delta_\text{g}$, and $^{6}\Sigma_\text{g}^+$ states, respectively \cite{Akash_2026_a}. Even higher-energy features beyond $\sim$33~eV were attributed to $^{8}\Sigma_\text{g}^-$, $^{8}\Pi_\text{g}$, $^{8}\Sigma_\text{g}^+$ and higher-lying states. A detailed discussion of the origin of these structures in connection to the different break-up mechanisms was presented in Ref.~\cite{Akash_2026_a}; here we focus on the differences among projectiles having the same charge $q$.

As can be seen in Fig.~\ref{figure1}, clear projectile-dependent differences emerge even when $q$ is held fixed, with the most pronounced contrast observed for \mbox{$q=4$}. While the KERDs for N$^{4+}$ and O$^{4+}$ projectiles are broadly similar, that for Ar$^{4+}$ differs markedly. For Ar$^{4+}$, the KERD is dominated by a single peak near 21~eV. In contrast, for N$^{4+}$ and O$^{4+}$, in addition to the peak at 21~eV, a broad and prominent maximum appears around 31~eV. This second peak coincides with the KER value predicted by the CE model. From our earlier calculations, the $^{6}\Sigma_\text{g}^+$ and $^{6}\Pi_\text{g}$ states are the most likely contributors to this feature \cite{Akash_2026_a}. Such a pronounced difference cannot be explained by classical charge-based models of electron capture alone. Within the quasi-molecular picture, where a transient (XCO$_2$)$^{q+}$ (X=$ $N, O and Ar) complex forms briefly during the collision, the observed differences reflect varying coupling strengths between entrance (X$^{q+}$ + CO$_2$) and exit channel (X$^{(q-3)+}$ + CO$_2^{3+}$) states, for a given final state of CO$_2^{3+}$. For Ar$^{4+}$ projectiles, the $^{2,4}\Pi_\text{g,u}$, $^{2,4}\Sigma_\text{g}^+$, and $^{2,4}\Delta_\text{g}$ states of CO$_2^{3+}$ in the exit channel appear to couple efficiently to the entrance channel, whereas for N$^{4+}$ and O$^{4+}$, the dominant coupling is instead to the $^{6}\Sigma_\text{g}^+$ and $^{6}\Pi_\text{g}$ states. Thus, even at the same charge and velocity, collisions involving N$^{4+}$ and O$^{4+}$ deposit energy into the target more strongly than Ar$^{4+}$ collisions.

For \mbox{$q=6$}, the KERDs for N$^{6+}$ and O$^{6+}$ projectiles are also broadly similar in shape. By contrast, Ar$^{6+}$ shows a noticeably stronger tendency to populate the high-KER shoulder above 25~eV. For \mbox{$q=8$}, the Ar$^{8+}$ KERD shows not only an enhanced shoulder but also a clear rightward shift relative to O$^{8+}$.
We remark that the KERDs obtained in the present work differ significantly from those reported earlier in slow \mbox{($v < 1$~a.u.)} \cite{Akash_2026_a,Akash_2026_b,Neumann_2010,Sumit_2021,Sumit_2022,Kumar_2024} and intermediate \mbox{($v \sim 1$~a.u.)} collision regimes \cite{Khan_2015,Yan_2016,Sumit_2022}. A detailed comparison with the earlier investigations is beyond the scope of the present work.

The strong similarity between the KERDs for N$^{q+}$ and O$^{q+}$ (\mbox{$q=4,6$}) and stark contrast of these two against the KERDs for Ar$^{q+}$, points to the importance of the binding-energy landscape of the post-collision projectile. This observation warrants a different test of projectile electronic structure effects, namely, comparing the KERDs with isoelectronic projectiles with unit difference in nuclear charge. Such comparison removes differences in electron count and occupancy from the problem and isolates changes in the binding energies and level spacings that govern capture.

\begin{figure}
	\centering
	\includegraphics[width=0.9\columnwidth]{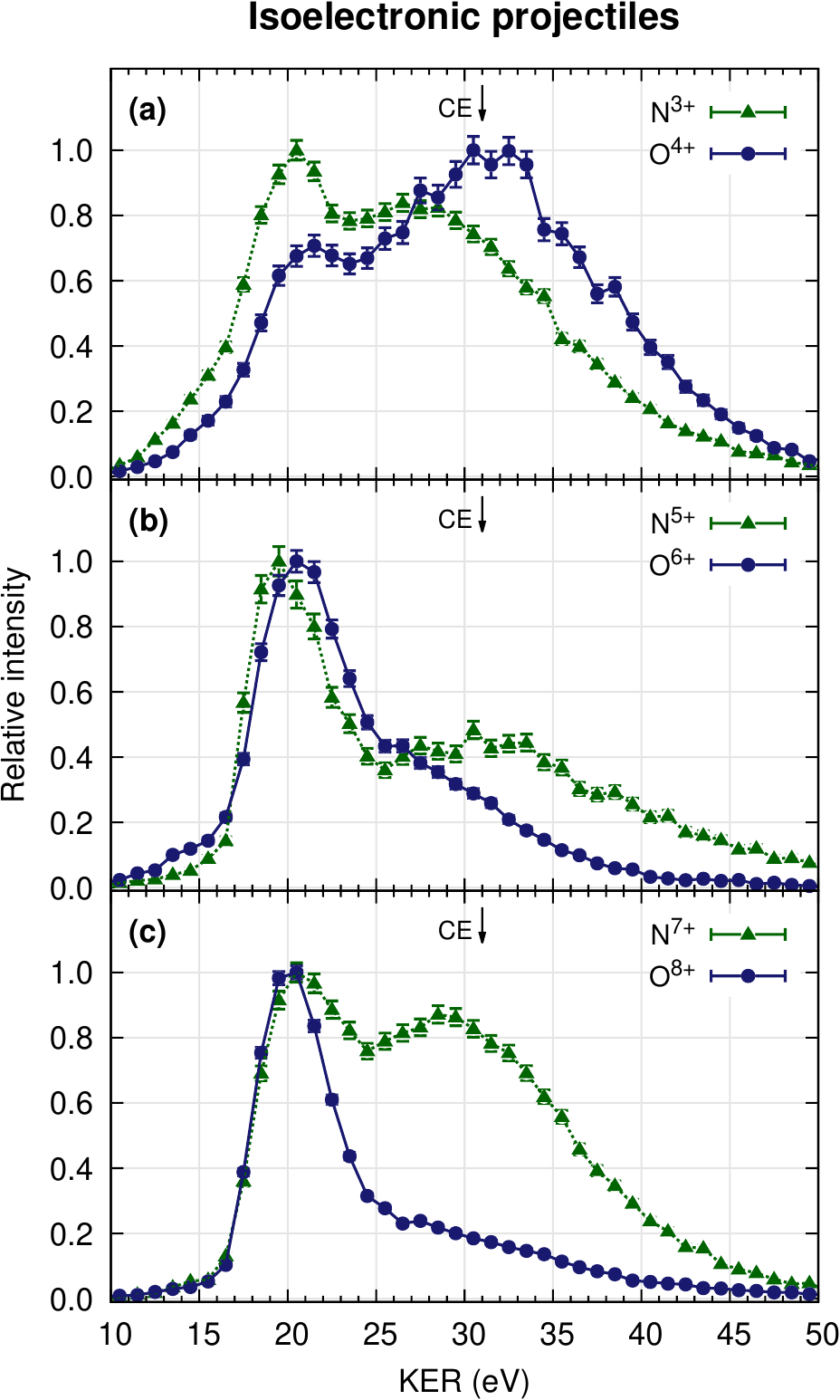}
	\caption{Same as Fig.~\ref{figure1}, but showing comparisons of the KERDs obtained with isoelectronic projectiles N$^{q+}$ and O$^{(q+1)+}$ in each panel.}
	\label{figure2}
\end{figure}

\begin{figure*}
	\centering
	\includegraphics[width=0.9\textwidth]{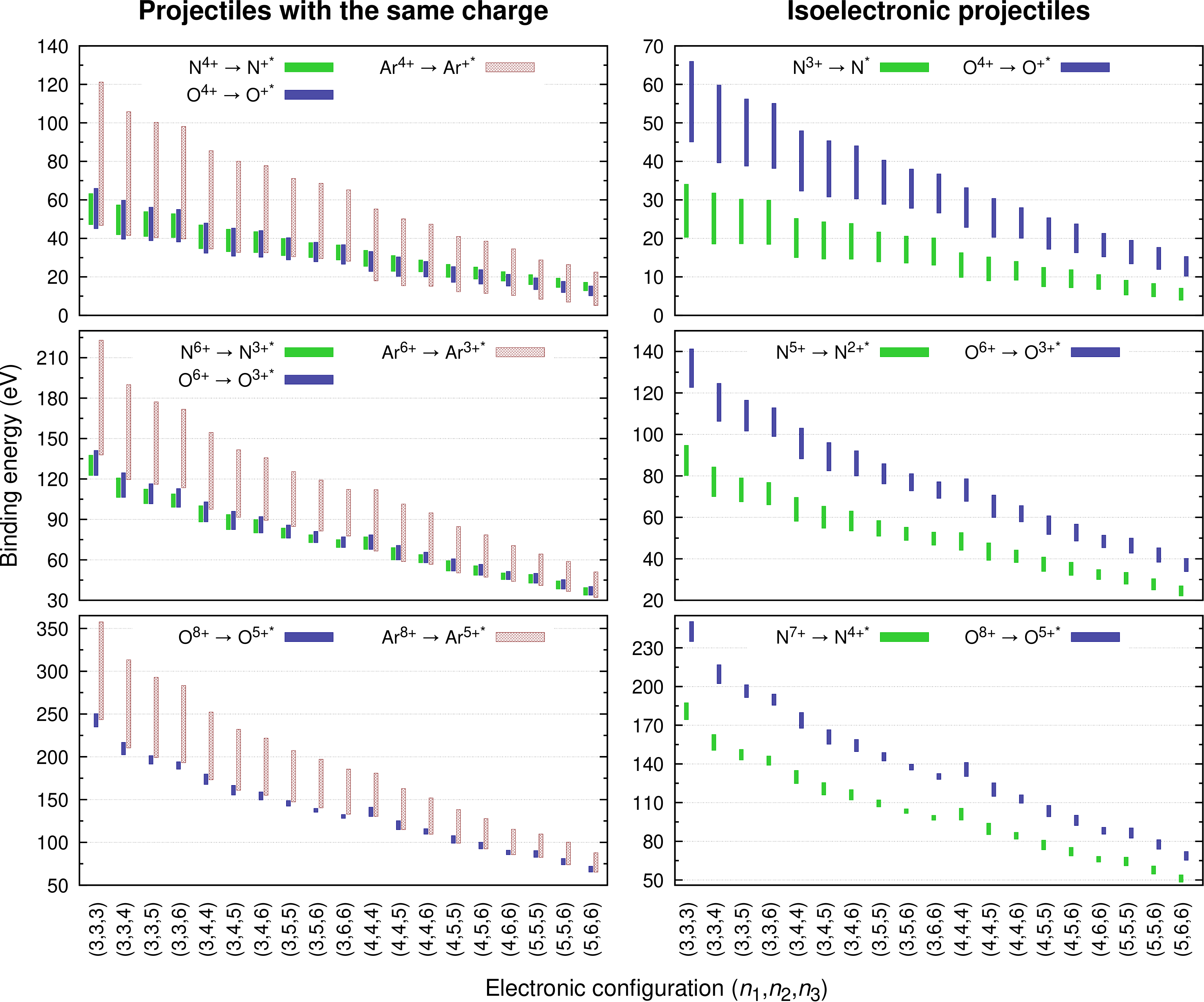}
	\caption{Binding energy ranges corresponding to various \mbox{($n_1l',n_2l'',n_3l'''$)} electronic configurations of the scattered projectile, labeled by their respective principal quantum numbers. The projectile is assumed to have undergone three-electron capture. Left panel shows the ranges for different projectile species with the same charge and the right panel shows the ranges for isoelectronic projectiles with unit difference in their charge.}
	\label{fig:BErange}
\end{figure*}

KERDs obtained with isoelectronic N$^{q+}$, O$^{(q+1)+}$ pair of projectiles are shown in Fig.~\ref{figure2}. A striking contrast is seen for the $q=3,4$ pair. KERDs for both N$^{3+}$ and O$^{4+}$ display a bimodal structure, but the relative intensities of the two features are reversed. A similar sensitivity is evident for the $q=5,6$ pair. The KERDs for N$^{5+}$ and O$^{6+}$ show nearly the same main peak, but the KERD for N$^{5+}$ has broad shoulder beyond 25~eV.  An even sharper contrast is observed for the bare projectiles N$^{7+}$ and O$^{8+}$. While the O$^{8+}$ KERD shows a single dominant peak, the N$^{7+}$ distribution is bimodal, with a secondary maximum near 29~eV.  We also see, that for O$^{q+}$ projectiles, increasing $q$ is accompanied by a systematic enhancement of the low-lying-state contribution, in line with the qualitative expectation from classical charge-scaling arguments. No such trend is evident for N$^{q+}$ and Ar$^{q+}$ projectiles, emphasizing that charge alone does not capture the full collision dynamics.  Within the framework of ECOBM and Bohr-Lindhard models, a unit increase in projectile charge would not be expected to induce substantial changes in the fragmentation energetics. However, pronounced differences are observed in the KERDs in each pair of projectiles, showing that even projectiles possessing the same electronic configuration but differing in charge by only one unit can lead to marked variations in the energy deposited to the target. That such differences persist even for bare projectiles (O$^{8+}$ and N$^{7+}$) highlights the point that the observed effects are not tied simply to the presence or absence of projectile electrons, but to the details of the energy landscape in the scattered projectile.

In light of these observations, it is pertinent to examine the binding energy (BE) ranges available in the scattered projectile. In the present velocity regime, CO$_2^{3+}$ is formed predominantly through capture of three electrons into excited states of the scattered X$^{(q-3)+*}$ projectile. To this end, we calculated the BE ranges for some of the relevant excited configurations using the flexible atomic code (FAC) \cite{Gu_2008}. These BE ranges are shown in \text{Fig.\ \ref{fig:BErange}} in two parts: for different projectile species with the same charge and for isoelectronic projectiles with unit difference in their charges. The main difference here is that in the former case, a much larger range of BE is available in the scattered Ar$^{(q-3)+*}$ projectiles, whereas the corresponding ranges in the scattered O$^{(q-3)+*}$ and N$^{(q-3)+*}$ are close to each other , while remaining substantially lower than those in Ar$^{q+}$. The similarity between the KERDs for N$^{q+}$ and O$^{q+}$ and their stark contrast against the KERDs for Ar$^{q+}$ go hand in hand with the BE patterns. However, when we go to the isoelectronic projectiles, the ranges of BE for a given pair are different and thus one may expect dissimilarity between the KERDs. However, there appears to be no facile relationship between the excitation of the target (which is what the KERDs reflect) and these BE ranges.  There are wide variations in the KERDs with the projectile species whereas, in contrast, the BE ranges show fairly monotonic variations.  This points to the need for a far more detailed, full quantum mechanical investigation of the interaction, beyond largely semi-classical descriptions.

Taken together, the present results provide direct evidence that multielectron capture cannot be understood in terms of projectile charge alone: the accessible post-collision states, and hence the coupling between entrance and exit channels, depend critically on the projectile electronic structure.

\acknowledgments
The authors thank the Dept.\ of Science and Technology, Science and Engineering Research Board (India) for generous funding via grant No. 30116294, which enabled the setting up of the EBIS/A ion source. They would also like to acknowledge technical help from Dreebit GmbH and assistance from the technical staff at IISER Pune in setting up and running the machine. A. S. acknowledges the award of fellowship from the Council of Scientific and Industrial Research (CSIR), India and financial support from Dr. V. S. Patankar Physics Endowment Fund, IISER Pune.

\bibliography{Projectile6}

\end{document}